\documentclass[aps,prl,showpacs,twocolumn]{revtex4}
\AtBeginDvi{}
\usepackage{graphicx}
\usepackage{epsfig}
\usepackage{dcolumn}% Align table columns on decimal point
\usepackage{bm}% bold math
\def\bra{\langle}
\def\ket{\rangle}
\begin{document}

\title{Anderson Lattice Description of Photoassociation in an Optical Lattice}

\date{\today}

\author{Takahiko Miyakawa and Pierre Meystre}
\affiliation{Department of Physics, University of Arizona, Tucson, Arizona 85721, USA}

\begin{abstract}
We consider atomic mixtures of bosons and two-component fermions
in an optical lattice potential. We show that if the bosons are in
a Mott-insulator state with precisely one atom per lattice, the
photoassociation of bosonic and fermionic atoms into heteronuclear
fermionic molecules is described by the Anderson Lattice Model. We
determine the ground state properties of an inhomogeneous version
of that model in the strong atom-molecule coupling regime,
including an additional harmonic trap potential. Various spatial
structures arise from the interplay between the atom-molecule
correlations and the confining potential. Perturbation theory with
respect to the tunneling coupling between fermionic atoms shows
that anti-ferromagnetic correlations develop around a spin-singlet
core of fermionic atoms and molecules.
\end{abstract}


\maketitle

% Introduction

Ultracold atoms and molecules trapped in optical lattices provide
an exciting new tool for the study of strongly correlated
many-body systems \cite{Jaksch05}. The exquisite degree of control
of the system parameters permits the detailed study of a variety
of exotic states of matter, and as a result these systems are
contributing to the establishment of significant new bridges and
interplay between AMO science and condensed matter physics. While
much work along these lines has concentrated so far on ultracold
atoms \cite{Jaksch98,Greiner02}, the coherent formation of bosonic
or fermionic molecules \cite{Rom04,Kohl05,AMLattice,Miyakawa04}
via either Feshbach resonances \cite{Inouye98} or two-photon Raman
photoassociation \cite{Wynar00} offers an additional path to the
study of strongly correlated atoms and molecules
\cite{Carr05,Dickerscheid05}.
Very recently, collective coherent phenomena between an atomic and
a molecular gas in an optical lattice have been observed experimentally
\cite{Heinzen05}.

In this letter we analyze the ground state of a mixture of atomic
bosons and two-component fermions coupled to heteronuclear
fermionic molecules~\cite{Stan04,Inouye04} by photoassociation or
Feshbach resonance in an optical lattice.
We show that this system can be mapped onto the Anderson Lattice
Model (ALM), a model that has previously found important
applications in the description of heavy electrons and
intermediate valence systems in condensed matter physics. In
particular, this model is known to exhibit a great variety of
possible behaviors, such as e.g. the Kondo effect and magnetic
ordering~\cite{Hewson}.

In the context of AMO experiments, the modifications of the
ground-state properties due to the presence of a trapping
potential are of particular interest. For example, in the bosonic
(fermionic) Hubbard model, that potential is known to result in
the coexistence of Mott-insulator and superfluid (metal)
phases~\cite{1DMSF,1DOLP}. In the strong atom-molecule coupling
regime under consideration here, the ALM exhibits two types of
magnetically correlated states of fermionic spins on the lattice,
an on-site spin-singlet (paramagnetic) state of the atoms and
molecules, and an anti-ferromagnetic (AF) correlated state among
lattice fermionic spins. The inhomogeneity of the confining
potential gives rise to a spatial structure of coexisting
paramagnetic and AF domains.

We consider at zero temperature a mixture of atomic bosons of mass
$M_b$ and atomic fermions of mass $M_a$ and spin $\sigma=\uparrow,
\downarrow$ trapped in an optical lattice potential~\cite{BFMlattice}.
The fermionic and bosonic atoms can be coherently combined into heteronuclear
fermionic molecules of mass $M_m=M_b+M_a$ via two-photon Raman
photoassociation. The lattice lasers are tuned so that the
fermionic atoms experience a weaker on-site lattice trapping
frequency, $\Omega_a$, than the bosonic atoms and the fermionic
molecules, $\Omega_b,\Omega_m\gg \Omega_a$.

The Hubbard-type Hamiltonian describing this system is
$\hat{H}=\sum_i (\hat{H}_{0i}+\hat{H}_{Ii})+\sum_{\langle ij
\rangle}\hat{W}_{ij}$, where
\begin{eqnarray}
\label{model}
\hat{H}_{0i}&=&\sum_{\sigma}\left\{\epsilon^a_{i\sigma}
\hat{a}_{i\sigma}^\dagger\hat{a}_{i\sigma}+(\epsilon^m_{i\sigma}+\nu)
\hat{m}_{i\sigma}^\dagger\hat{m}_{i\sigma}\right\}
+\epsilon^b_i\hat{b}_{i}^\dagger\hat{b}_{i}\nonumber\\
&+&U_{a}\hat{n}^a_{i\uparrow}\hat{n}^a_{i\downarrow}
+U_{m}\hat{n}^m_{i\uparrow}\hat{n}^m_{i\downarrow}+\frac{1}{2}U_{b}
\hat{n}^b_i(\hat{n}^b_i-1),\\
\hat{H}_{Ii}&=&g\sum_{\sigma}(\hat{m}_{i\sigma}^\dagger
\hat{b}_{i}\hat{a}_{i\sigma} +{\rm H.c.}),\\
\hat{W}_{ij}&=&\sum_\sigma \left(-t_a\hat{a}_{i\sigma}^\dagger
\hat{a}_{j\sigma}-t_m\hat{m}_{i\sigma}^\dagger \hat{m}_{j\sigma}\right)
-t_b\hat{b}_{i}^\dagger \hat{b}_{j}+{\rm H.c.}\quad
\end{eqnarray}
Here $\hat{a}_{i\sigma}$, $\hat{m}_{i\sigma}$, and $\hat{b}_i$ are
the annihilation operators of the fermionic atoms and molecules
and of the bosonic atoms at the $i$-th site, respectively. The
corresponding number operators $\hat{n}^a_{i\sigma}
=\hat{a}_{i\sigma}^\dagger\hat{a}_{i\sigma}$,
$\hat{n}^m_{i\sigma}=\hat{m}_{i\sigma}^\dagger\hat{m}_{i\sigma}$,
and $\hat{n}^b_{i}=\hat{b}_{i}^\dagger\hat{b}_i$ have eigenvalues
$n^a_{i\sigma}$, $n^m_{i\sigma}$, and $n^b_{i}$.

The on-site single-particle energies $\epsilon^\alpha_{i\sigma}$,
where $\alpha=a,b,m$, consist of the sum of the contributions
$\epsilon^\alpha \approx \Omega_\alpha$ from the lattice and
$V^\alpha_{i,\sigma}$ from the trap potential at site $i$, the
two-photon detuning between the Raman-lasers and the difference in
internal energies between the atoms and molecules being included
in $\nu$. In case the external potential is created by optical
means, the spin structure of the fermions is not resolved,
resulting in spin-independent single particle energies. This is
the case that we consider here. The terms proportional to $U_a$,
$U_m$, and $U_b$ describe on-site two-body interactions,
inter-species collisions being ignored for simplicity. Finally,
the interaction Hamiltonians $\hat{H}_{Ii}$ describes the
conversion of fermionic and bosonic atoms into fermionic
molecules, and tunneling between nearest neighbor sites denoted by
$\langle ij \rangle$ is described by the parameters $t_a$, $t_m$,
and $t_b$.

We concentrate on the case $U_{b,m}\gg g\gg t_{b,m}$ and $g \agt
t_a\gg U_a$, a regime that is achieved for $\epsilon^{b,m}\gg
\epsilon^{a}$ and $M_b\gg M_a$. One can then ignore intersite
tunneling of the fermionic molecules and bosonic atoms, whose rate is
given by the exponential of the ratio between the lattice constant
$\lambda/2$ and the width of the localized state
$l_\alpha=(\hbar^2/M_\alpha \epsilon_\alpha)^{1/2}$
\cite{Dickerscheid05}. We assume the opposite condition for the
fermionic atoms and neglect instead their on-site two-body
collisions~\footnote{A gas of the two lowest hyperfine states of $^6$Li is a possible candidate, where the scattering length of the mixture is very small at zero magnetic field. See, e.g., Ref.~\cite{OHara02}}. Typical parameters for $^{87}$Rb atoms are
$\epsilon_b/\hbar\simeq 10^5$ s$^{-1}$ and $U_b/\hbar\simeq 10^4$
s$^{-1}$, and the photoassociation coupling constant is estimated
to be of the order of $g/\hbar\simeq 10^3$
s$^{-1}$~\cite{Miyakawa04}. Since heteronuclear diatomic molecules
are polar, we also need an estimate of their electric
dipole-dipole interaction. The dipole moment of the triplet
$^3\Sigma^+$ state of the KRb dimer is $0.02$ a.u.
~\cite{Kotochigova03}, ($1$ a.u. is $8.478\times 10^{-30}$ Cm), in
which case the dipole-dipole interaction energy is less than $10$
s$^{-1}$ for $\lambda=1064$ nm and can safely be ignored. We
further neglect all loss processes.

In the absence of photoassociation, the ground state of the atomic
fermions consists of free fermions in the confining trap $V^a_i$
while for bosons it is a Mott-insulator core with exactly one
boson per site within a set $\{\ell\}$ of sites from the center of
trap up to edges determined by the number of bosons $N_b$, and
zero occupation outside. We therefore have
\begin{equation}
\label{constraint}
\sum_{\sigma}\hat{n}^m_{i\sigma}+\hat{n}^b_i=1\; {\rm for} \,i\in
\{\ell\}; \,\,\,\,\ \sum_{i\sigma}\hat{n}^f_{i\sigma}=N_f,
%\sum_{i\sigma}\hat{n}^m_{i\sigma}+\sum_{i\sigma}\hat{n}^a_{i\sigma}=N_f.
\end{equation}
where
$\hat{n}^f_{i\sigma}=\hat{n}^a_{i\sigma}+\hat{n}^m_{i\sigma}$.
The former constraint is due to prohibition of double occupations
and neglect of tunnelings for bosonic atoms and fermionic molecules~\footnote{More rigorously, this local constraint holds in the case that the time scales of $1/t_b$ and $1/t_m$ are much longer than that of change of photoassociation process. In contrast, the change is assumed to be adiabatic for any other time scales including atomic tunneling $1/t_a$.}.
These restrictions result in the reduced Hamiltonian
\begin{eqnarray}
\label{ALM}
\hat{H}_{r}&=&-t_a\sum_{\langle ij \rangle,\sigma}(\hat{a}_{i\sigma}^\dagger
\hat{a}_{j\sigma}+{\rm H.c.})+\sum_{i\sigma} \epsilon^a_i
\hat{a}_{i\sigma}^\dagger\hat{a}_{i\sigma}\\
&+&\sum_{i\sigma} (\nu+\epsilon^m_i-\epsilon^b_i)\hat{m}_{i\sigma}^\dagger
\hat{m}_{i\sigma}
+g\sum_{i\sigma} (\hat{m}_{i\sigma}^\dagger \hat{b}_{i}
\hat{a}_{i\sigma} +{\rm H.c.}).\nonumber
\end{eqnarray}
In the homogeneous case $V^\alpha_i=0$, this Hamiltonian is
equivalent to the so-called {\it slave boson model}, which under
the constraints~(\ref{constraint}) reduces to the periodic ALM
with infinite on-site repulsion. We note, however, that in our
model the bosonic operator $\hat{b}$ corresponds to a real
physical particle, in contrast to the pseudo-boson introduced in
the study of the Kondo problem~\cite{Hewson}. In the following
we investigate the ground state properties of this system in the
presence of a harmonic trap and for weak tunneling of the
fermionic atoms. Since the atom-molecule (AM) correlation energy is
order of $g$ as shown later, this is justified in the regime $g\gg
t_a$. We proceed by first applying a counting argument for the
case $t_a=0$ under the constraints~(\ref{constraint}), and then account for tunneling
coupling in perturbation theory.

For $t_a=0$, the ground state is a product of lowest energy states
with total fermion number $n^f=n^a+n^m=\sum_\sigma
n^a_{i\sigma}+\sum_{\sigma} n^m_{i\sigma}$ on each lattice site,
%\begin{equation}
$|\Phi_0\rangle=\prod_i|n^f;{\rm lowest \; energy}\rangle_i$.
%\end{equation}
For lattice sites belonging to the set $\{\ell\}$, we can have
$n_f = 0,\ldots,3$ fermions, and the corresponding normalized
lowest energy states are
    \begin{eqnarray}
    &&|n^f=0\rangle_l = |0_a,0_m\rangle_l, \label{uncorr0} \\
    \label{corr1}
    &&|n^f=1;\sigma\rangle_l = \alpha_l|\sigma_a,0_m\rangle_l
    +\beta_l|0_a,\sigma_m\rangle_l, \\
    \label{corr2}
    &&|n^f=2\rangle_l = u_l|2_a,0_m\rangle_l
    +v_l\left(|\uparrow_a,\downarrow_m\rangle_l-|\downarrow_a,\uparrow_m\rangle_l
    \right),\\
    \label{uncorr3}
    &&|n^f=3;\sigma\rangle_l = |2_a,\sigma_m\rangle_l,\quad
    \end{eqnarray}
with corresponding energies
\begin{eqnarray}
E_l(0)&=&0,\nonumber\\
E_l(1)&=&\epsilon^a_l+\delta_l-\sqrt{\delta_l^2+g^2},\nonumber \\
E_l(2)&=&2\epsilon^a_l+\delta_l-\sqrt{\delta_l^2+2g^2},\nonumber \\
E_l(3)&=&3\epsilon^a_l+2\delta_l,
\end{eqnarray}
where we have used the compact notations
$|n\sigma_a,n^\prime\sigma^\prime_m\rangle_l=
|n\sigma_a\rangle_l\otimes|n^\prime\sigma^\prime_m\rangle_l
\otimes|n^b=1-n^\prime\rangle_l$ and
$2\delta_l=\nu+\epsilon^m_l-\epsilon^b_l-\epsilon^a_l$. Here
$\alpha_l=-g\{[(\delta^2_l+g^2)^{1/2}-\delta_l]^2+g^2\}^{-1/2}$
and
$u_l=-\sqrt{2}g\{[(\delta^2_l+2g^2)^{1/2}-\delta_l]^2+2g^2\}^{-1/2}$.
The fermionic atom and molecule populations at a single site are
$\langle n^a_l\rangle=\alpha_l^2$ and $\langle
n^m_l\rangle=\beta_l^2$ respectively for the $n^f=1$ manifold, and
$\langle n^a_l\rangle=1+u^2_l\leq 2$ and $\langle n^m_l
\rangle=2v_l^2 \leq 1$ for $n^f=2$. We observe that the average
number of fermionic atoms gradually decreases with decreasing
$\delta_l$, while the number of molecules increases. From Eqs.
(\ref{corr1}) and (\ref{corr2}), we also note that the
atom-molecule conversion leads to correlated states in the $n^f=1$
and $n^f=2$ manifolds, with energies lowered by an amount of the
order of $g$. The state $|n^f=2\rangle$ corresponds to a spin-singlet state,
while spin-triplet states such as ferromagnetic configuration
have no coherent scattering and then are higher energy states.

To count the number of fermions and find the ground state
configuration, we introduce at each lattice site $l$ the local
chemical potential $\eta_l(M)\equiv E_l(M)-E_l(M-1)$ ($M=1,2,3$).
It corresponds to the energy needed to add one fermion to $M-1$
fermions already at that site. At the sites $i\notin \{\ell\}$,
the local chemical potential is equivalent to the atomic single
particle energy, $\eta^1_i=\eta^2_i=\epsilon^a_i$. We then proceed
by first determining the Fermi energy $\epsilon_F$ for a
fixed number of fermions, the fermionic occupation number $n^f$ at
a given site being then determined by the condition
$\eta_l(n^f)\leq\epsilon_F\leq \eta_l(n^f+1)$, where we set
$\eta_l(0)\to -\infty$ and $\eta_l(4)\to +\infty$ for convenience.

%This procedure yields for each site the phase diagram of $n^f$ as
%function of $\delta_l/g$ and $(\epsilon_F-\epsilon^a_l)/g$ shown
%in Figure~\ref{fig0}. The gray shading indicates the degree of the
%atom-molecule joint coherence, $j_l\equiv \sum_{\sigma} {\it Re}
%\langle \hat{m}^\dagger_{l\sigma}
%\hat{b}_l\hat{a}_{l\sigma}\rangle$. The atom-molecule coherence in
%the $n^f=1$ and $n_f=2$ manifolds is maximum at $\delta=0$. As
%evident from Eqs. (\ref{uncorr0}) and (\ref{uncorr3}) there are no
%such coherence in the $n^f=0$ and $n^f=3$ manifolds. Finite
%tunneling results in a smearing of the borders between different
%phases.
%
%\begin{figure}
%  \begin{center}
%    \includegraphics[width=6cm,height=4cm]{fig0}
%  \end{center}
%  \caption{Phase diagram of $n^f$ at the $l$-th site
%    as functions of $\delta_l/g$ and $(\epsilon_F-\epsilon^a_l)/g$.
%  Shading describes the degree of the atom-molecule joint coherence $j_l$.}
%  \label{fig0}
%\end{figure}
%

In order to gain insight into the ground-state properties
of the system, we consider the specific case of a one-dimensional,
harmonic trap potential $V^{\alpha}_i$ for each species
$\alpha=a,b,m$~\cite{1DOLP},
%\begin{equation}
$V^\alpha_i=V^\alpha\{2/(N_b-1)\}^2\{i-(N_b+1)/2\}^2$,
%\end{equation}
with the trap center located half-way between two lattice sites.
%The extension of this potential to higher dimensions is straightforward. 
We assume that the strength of molecule trapping
is the sum of those of the atomic fermions and bosons, i.e.,
$V^m=V^a+V^b$, resulting in the uniform detuning
$2\delta_l=2\delta\equiv\nu+\epsilon^m-\epsilon^a-\epsilon^b$. We
also set the offset energy $\epsilon^a=0$ without loss of
generality.

\begin{figure}
  \begin{center}
    \includegraphics[width=7.5cm,height=6cm]{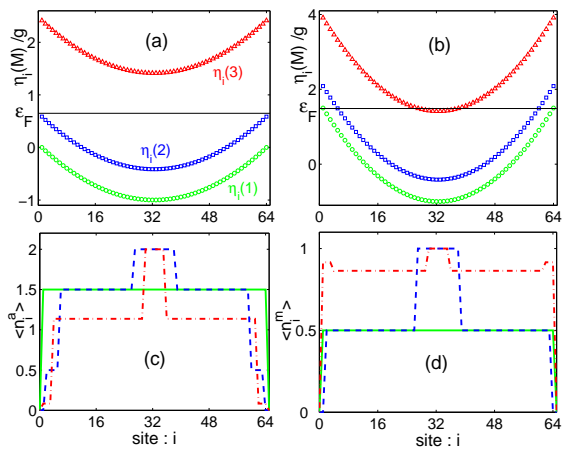}
  \end{center}
  \caption{(Color online) $\eta_i(M)$ for $\delta/g=0$ and $N_b=64$,
   and for (a) $V^{a}/g=1$, (b) $V^{a}/g=2.5$; Ground state density
   distribution of (c) fermionic atoms, (d) molecules
    for $\delta/g=0$, $V^a/g=1$ (solid line), $\delta/g=0$,
    $V^a/g=2.5$ (dashed line), and $\delta/g=-1.5$, $V^a/g=1$
    (dot-dashed line) for $N_f=2N_b=128$.}
  \label{fig1}
\end{figure}

Figures~\ref{fig1}a,b show the local chemical potentials
$\eta_l(M)$ at each lattice site for $\delta=0$ and for (a)
$V^a/g=1$ and (b) $V^a/g=2.5$. 
In this example, the system would contain
$64$ bosonic atoms occupying lattice sites within $1\leq l \leq
64$ in the absence of photoassociation, $g=0$. We remark that in
that limit all $\eta_l(M)$ become
degenerate and equal to $\epsilon^a_l$ at $\delta=0$,
clearly indicating that the gaps among the three chemical
potential ``bands'' arise from the AM coupling.
The corresponding Fermi energies for $N_f=2N_b=128$  are the solid lines. 
In case (a), the ground state corresponds to a
configuration where all sites within $\{\ell\}$ are in the 
spin-singlet state $|n^f=2\rangle$, all other sites being in the
$|n^f=0\rangle$ state. In contrast, in case (b) the trap energy
near the edges of the atomic density overcomes 
the AM correlation energy at the center of trap. As a
result, the fermions are in the uncorrelated state $|n^f=3\rangle$
near the center of the trap and in the correlated state
$|n^f=1\rangle$ at the edges.

The ground-state density distributions of fermionic atoms $\langle
\hat{n}^a_i\rangle$ and molecules $\langle \hat{n}^m_i\rangle$ are shown in
Fig.~\ref{fig1}c and \ref{fig1}d, respectively. For $V^a/g=1$ (solid line),
spatially homogeneous densities of fermions are realized up to the
edge of the sites $\{l\}$ and form the spin-singlet state of
Eq.~(\ref{corr2}). This corresponds to the {\it Kondo
insulator}~\cite{Hewson} state in homogeneous systems. Because of
the correlated gap energy this state is stable against small
amounts of atomic tunneling $t_a$. For $V^a/g=2.5$ (dashed line),
atoms and molecules form a spatial shell structure reminiscent of
the Mott-insulator shells of bosons or fermions 
%resulting from the interplay between the confining potential and the on-site repulsive interaction 
in combined optical lattices and trapping potentials~\cite{1DMSF,1DOLP}.
As $\delta/g$ becomes large and
negative, the molecule population at each lattice site approaches
unity, but the inhomogeneous character of the population persists
even for $V^a/g=1$ at $\delta/g=-1.5$ as shown by the dot-dashed
lines in Fig.~\ref{fig1}c,d.

We now discuss the effects of weak intersite tunneling of the
fermionic atoms. The role of the tunneling is two folds.
First, the enhancement of fluctuations of fermion number
on the border between different $n^f$ shells.
Second, the anti-ferromagnetic (AF) correlation on the 
$|n^f=1;\sigma\rangle$ domain. In the strong AM coupling regime,
$g\gg t_a$ and $|\delta|\lesssim g$, these can be investigated by
(quasi-)degenerate perturbation theory up to second order in $t_a$.
The ground state is given by
$$
|\Psi_g\rangle=\sum_{\{n^f_i,\sigma_i\}}C_g{\{n^f_i,\sigma_i\}}
|\Phi_0\{n^f_i,\sigma_i\}\rangle.
$$
The set of $\{n^f_i,\sigma_i\}$ contains those fermion number 
configurations, $\{n^f_i\}$, which can be realized
through multiple tunneling with an only energy cost $|V^a_j-V^a_i| (\ll g)$ 
for each process from the unperturbed ground state in the absence of $t_a$. 
A possible spin configuration $\{\sigma_i\}$ can couple to a different one
via virtual excitations of energy lager by an amount of order $g$.

\begin{figure}
  \begin{center}
    \includegraphics[width=8.5cm,height=4cm]{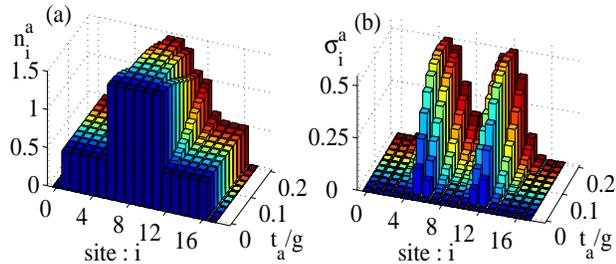}
  \end{center}
  \caption{(Color online) (a) $\langle \hat{n}^a_i\rangle$ and
(b) $\sigma^a_i$ as a function of $t_a/g$
for $N_b=16$, $N_f=22$ and for $\delta/g=0$, $V^a/g=1.5$.}
  \label{fig2}
\end{figure}

To illustrate the role of the number fluctuations we consider the case
$N_b=16$, $V/g=1.5$, and $\delta=0$ with $N_f=22$ such that the unperturbed
ground state in the system of $t_a=0$ with $|n^f=1\rangle$
domains surrounding the spin-singlet core of $|n^f=2\rangle$
and no $|n^f=3\rangle$ component.
Figure~\ref{fig2} shows the density and number fluctuations of fermionic atoms,
defined by $(\sigma^a_i)^2=(\langle (\hat{n}^a_i)^2 \rangle - \langle \hat{n}^a_i \rangle^2)/ \langle \hat{n}^a_i \rangle$,
as a function of $t_a/g$ for the direct diagonalization in 
the first-order perturbation theory.
With increasing $t_a$ the hybridization of the different $n^f$ states
is higher on the border between $n^f=1$ and $n^f=2$ shells,
while the spin-singlet core and $|n^f=1\rangle$ domains remain.
Consistently with this result, the number fluctuations of atoms develop 
strongly on around the border region and weakly on the core and 
surrounding regions.
The reason of the suppression of fluctuations on the edge 
of the sites $\{l\}$ is that
atomic tunneling from the edge sites into outside costs
an energy $\sim g$, so that such a transition is neglected.
We note that the result of Fig.~\ref{fig2} does not depend on qualitatively
the excat value of the detuning parmeter and number fluctuations of molecules
are vanishingly small for any $|\delta| \lesssim g$ as expected.

The second-order corrections of atomic tunneling between nearest neighbor sites that are both within the $|n^f=1\rangle$ manifold leads to an AF correlation among fermionic spins composed of mixtures of atoms and molecules.
In contrast, if at least one site is in a $n^f \neq 1$ manifold, those tunneling processes are spin-independent and result only in an energy shift.
The origin of AF correlation is similar to that observed in Hubbard model
in the strong coupling limit~\cite{Nagaosa}, while in our problem at hand
the coherence character between atoms and molecules plays a key role.
%in this mechanism.
%While spin-flip can only occur between anti-parallel neighboring atomic spins, 
%it is possible between a site occupied by an atom and a neighboring one occupied by a molecule because of superposition state of $|n^f=1\ranlge$.
%Hence the AF correlation is seen to result from
%the combined effects of a suppression of intersite tunneling
%involving atoms of parallel spins.
%together with the lowered
%energy of unperturbed excited states with anti-parallel fermionic spins 
%in a consequence of the AM coherence.
%

\begin{figure}
  \begin{center}
    \includegraphics[width=6.cm,height=3.5cm]{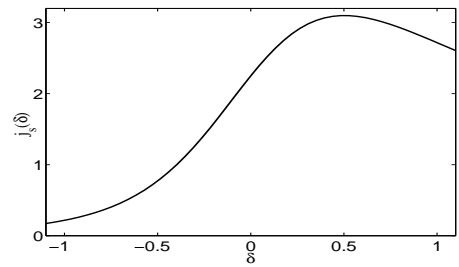}
  \end{center}
  \caption{Scaled AF coupling strength $j_s$ as a function of $\delta$.}
  \label{fig3}
\end{figure}

Introducing the coherent spin operators at the $i$-th site
$s^f_{i,z}=(1/2)(\hat{n}^f_{i\uparrow}-\hat{n}^f_{i\downarrow})$,
$s^f_{i,+}=(s^f_{i,-})^\dagger=\hat{a}^\dagger_{i\uparrow}\hat{a}_{i\downarrow}
+\hat{m}^\dagger_{i\uparrow}\hat{m}_{i\downarrow}$$,$,
the spin-dependent effective Hamiltonian is
\begin{equation}
\label{effectiveH}
\hat{H}_s=\sum_{\langle ij \rangle \in  n^f=1}J_{ij}
\left(\bm s^f_i \cdot \bm s^f_j -\frac{1}{4}\right)
-\frac{t^2_a \alpha^2 \beta^2}{\sqrt{\delta^2+g^2}}.
\end{equation}
As a consequence of AM coherence, the AF coupling coefficient
$J_{ij}(>0)$ depends on the detuning parameter and,
when $|V^a_j-V^a_i|\ll g$, $J_{ij}$ is reduced to a site-independent one
$J=t^2_a/g \cdot j_s(\delta)$ shown in Fig.~\ref{fig3}.
The last term of the above Hamiltonian is owing to possible tunneling between a site occupied by an atom and a neighboring one occupied by a molecule 
even for parallel spins.

Since Hamiltonian (\ref{effectiveH}) is
spin $1/2$ Heisenberg model, we can make out it's magnetic properties
on a $|n^f=1\rangle$ domain.
If we consider a one-dimensional example like Fig.~\ref{fig2},
Hamiltonian~(\ref{effectiveH}) affects left and right sides
separated by spin-singlet core independently.
The ground state corresponding to each domain is characterized by
$\sum_{i=1}^{L_D} s^a_{iz}=0$. The asymptotic form of
the spin density correlation function has
a power law behavior~\cite{Nagaosa}
$
\bra s^f_{j+k,z} s^f_{j,z}\ket \sim (-1)^k/k
$
except for finite size corrections.

The present mechanism of AF correlations is unique in that atomic
tunneling and strong AM couplings act cooperatively,
resulting in coehrent AF characteristics among intersite fermionic spins of
both atoms and molecules. Moreover, the AF correlation develops outside
the paramagnetic spin-singlet core, so that the spatial paramagnetic
and AF correlated domains coexist.
For finding  magnetic properties on the border region
or when $J \sim t_a \ll g$, we need to consider the interplay
of number fluctuations and AF correlations.
This will be discussed in a future publication.

In summary, we have studied atomic boson-fermion mixtures with
atom-molecule conversion in the presence of an optical lattice
potential plus a harmonic oscillator potential. We have shown that
by controlling lattice depths for fermions and bosons
independently, an inhomogeneous version of the ALM can be
realized. The new feature of inhomogeneity gives rise to various
spatial structures of correlated atom-molecule ground states. We
have analyzed the tunneling coupling effect by degenerate
perturbation theory and have shown that this coupling and
inhomogeneity give rise to coexistence of spatial domains of
paramagnetic and AF correlations. 
%Since the AF ordering might be
%possible in higher dimensions, future work will explore the
%possible existence of an exotic magnetic ordered ground state and
%collective excitations in this inhomogeneous system.

\end{document}